\documentclass[hyper]{article}
\usepackage{jheppub}
 \usepackage{enumerate}
 \usepackage{url}
\input{epsf}
\usepackage{epsfig}
\usepackage[all]{xy}
\usepackage{amsmath,amsthm,simpler-wick}
\usepackage[mathscr]{eucal}
\usepackage{makecell,mathtools,physics}
\newcommand\addtag{\refstepcounter{equation}\tag{\theequation}}

\newcommand{\bbN}{{\mathbb{N}}}

\newcommand{\bbZ}{{\mathbb{Z}}}

\newcommand{\mcC}{{\mathcal{C}}}

\newcommand{\mcN}{{\mathcal{N}}}
\newcommand{\mcO}{{\mathcal{O}}}

\newcommand{\mcY}{{\mathcal{Y}}}

\DeclareMathOperator{\U}{U}
\DeclareMathOperator{\SO}{SO}
\DeclareMathOperator{\SU}{SU}
\DeclareMathOperator{\SL}{SL}

\newcommand{\normord}[1]{~\xcentcolon\mathrel{#1}\xcentcolon~}
\newcommand{\xcentcolon}{%
  \mathrel{\vbox{\hbox{$:$}\kern.1ex}}%
}

\newcommand{\IZ}{{\mathbb{Z}}}
\newcommand{\IR}{{\mathbb{R}}}
\newcommand{\sgn}{\text{sgn}}

\title{One-loop Renormalization of BPS String Masses in Pseudo-anomalous Heterotic String }

\author[1]{Jeffrey A. Harvey,}
\author[1]{Tai Wai Hu}

\affiliation[1]{Kadanoff Center, Enrico Fermi Institute and Department of Physics\\
$~~ $University of Chicago \\
$~~ $5620 Ellis Ave., Chicago IL 60637}

\emailAdd{j-harvey@uchicago.edu}
\emailAdd{twhu@uchicago.edu}

\abstract{
Compactification of heterotic string on a Calabi-Yau threefold can lead to a four-dimensional low-energy effective theory
which contains a $\U(1)$ gauge theory which is pseudo-anomalous, meaning that the fermion content is anomalous, but that the fermion anomaly is cancelled by a four-dimensional version of the Green-Schwarz mechanism involving a shift of the model-independent axion field.   
It is also well known that a field-dependent Fayet-Iliopoulos like term is induced at one-loop in string perturbation theory in such compactifications and that this leads to a mass for the $\U(1)$ gauge field. Explicit one-loop computations of the mass shifts of massless charged scalars and fermions
implied by this mechanism were originally carried out in the 1980's and have been revisited more recently using techniques of string field theory and super Riemann surfaces. We consider such theories further
compactified on a circle of radius $R$. There is then an infinite tower of BPS states with winding and momenta on the circle. BPS strings which wind the circle can, at large $R$,  be viewed as macroscopic or cosmic strings in four spacetime dimensions, or, at generic $R$, as BPS states in three dimensions. We compute the one-loop mass renormalization of such states and show that it is nonzero and proportional to their pseudo-anomalous $\U(1)$ charge.

\vskip 0,1in
\today}


\begin{document}
\maketitle

\addtocontents{toc}{\protect\enlargethispage{35mm}}

\section{Introduction}\label{sec:Intro}

This paper involves a combination of two topics in string theory which date back to the late 1980's, namely the study of Fayet-Iliopoulos like terms\footnote{Fayet-Iliopoulos terms have a long and complicated history.
As discussed  in \cite{Komargodski:2009pc} such terms with constant coefficients cannot be consistently coupled to supergravity. However FI-like terms, but with field dependent coefficients are consistent and arise in string compactifications. The terminology ``field-dependent FI terms" was discouraged in \cite{Komargodski:2009pc}, but no alternative
name was proposed, so we have adopted the not entirely satisfactory nomenclature FI-like terms.} 
in heterotic string compactifications \cite{Dine:1987xk, Atick:1987gy, Atick:1987qy,Dine:1987gj, Dine:1989gb}  and the discovery that supersymmetric heterotic string compactifications with an $S^1$ factor contain an open string spectrum of perturbative BPS states \cite{Dabholkar:1989jt, Dabholkar:1990yf}. 

Many compactifications of string theory lead to the presence of a $\U(1)$ gauge theory with anomalous fermion content.  The first example in the literature \cite{Dine:1987xk}, and one that will play a central role in this paper, arises in the study of the $\SO(32)$ heterotic string compactified on a Calabi-Yau threefold $\mcY$ with the standard embedding
of the $\SU(3)$-valued spin connection on $\mcY$ into the $\SO(32)$ gauge group. This breaks $\SO(32)$ to $\SO(26) \times \U(1)$, and when the Euler number $\chi(\mcY)= 2(h^{1,1}-h^{2,1})$ is non-zero the fermion content in the effective theory is anomalous. There are $\U(1)$ gravitational anomalies proportional to the trace of the $\U(1)$ generator $Q$ over the left-handed
massless fermion fields, and $\U(1)^3$ and $\U(1)-\SO(26)-\SO(26)$ gauge anomalies. These anomalies are all canceled by a four-dimensional version of the Green-Schwarz
mechanism which in this context involves a shift in the model-independent axion field $a$ \cite{Dine:1987xk}.

 In these theories there is a term that plays an important role
in the low-energy effective action  which was called a Fayet-Iliopoulos (FI) D-term in earlier literature. FI terms in superspace take the form
\begin{equation}
{\cal L}_{\text{FI}}=c \int d^4 \theta V
\end{equation}
where $V$ is the $\U(1)$ vector superfield, and arise in the following way. In the low-energy description of the string compactification we study one finds a modification of the Kahler potential for the the axion-dilaton multiplet $S$ (whose scalar component $s_{R} + ia$ combines the axion with the dilaton $s_{R}$) of the form
\begin{equation}
K=-M_P^2 \log (S+S^* + c V)
\end{equation}
and this leads to terms which resemble the FI term, but multiplied by terms depending on the dilaton field. After
eliminating the auxiliary field in the $V$ multiplet one finds a potential for charged scalar fields $\phi_i$ which, when the dilaton
takes a constant value $s_R=g_s^{-2}$, and after rescaling to string theory rather than supergravity conventions, has the form 
\begin{equation}
g_s^{-2} \left(\sum_i q_i \phi_i^* \phi_i - c M_s^2 g_s^2 \right)^2= g_s^{-2} \qty(\sum_i q_i \phi_i^* \phi_i)^2 - 2 c M_s^2 \sum_i q_i \phi_i^* \phi_i + c^2 M_s^4 g_s^2 
\end{equation}
where $q_i$ is the $\U(1)$ charge of $\phi_i$ and $M_s$ is the string scale. 
Recalling the string loop expansion is a genus expansion with genus $g$ world-sheets contributing effects of order $g_s^{2g-2}$, we see that 
the coefficient $c$ which leads to a mass term for the $\phi_i$ appears first at genus-one. It can therefore be determined either from a one-loop computation of the correction to the mass of any of the $\phi_i$ \cite{Atick:1987gy, Dine:1987gj} or from a contribution to the two-loop dilaton tadpole \cite{Atick:1987qy}.

In these theories the gauge field of the pseudo-anomalous $\U(1)$ acquires a mass due to two effects. First, under $\U(1)$ gauge
transformations $A_\mu \rightarrow A_\mu + \partial_\mu \Lambda$ the axion undergoes a shift symmetry $a \rightarrow a+ \beta \Lambda$
for some constant $\beta$. This shift when combined with couplings of the form $a {\rm Tr} F \wedge F$ and $a {\rm Tr} R \wedge R$ cancels
the anomalies coming from the chiral fermions. The kinetic term for $a$ then takes the form $(\partial_\mu a - \beta A_\mu)^2$
which exhibits $a$ as part of the longitudinal degree of freedom of $A_\mu$.  Second, it is generically the case that there are charged fields $\phi_i$ which can acquire
an expectation value which leads to vanishing vacuum energy and a new vacuum with unbroken supersymmetry. For example, in the
$\SO(32)$ model one of the charged $\SO(26)$ singlet scalars can acquire an expectation value. This also breaks the pseudo-anomalous $\U(1)$
symmetry. As a result, the longitudinal component of the massive $\U(1)$ gauge bosons is a linear combination of the original axion $a$ and
a component of the $\phi_i$ acquiring an expectation value while the orthogonal linear combination remains massless and has axion like
couplings to ${\rm Tr} F \wedge F$ and ${\rm Tr} R \wedge R$.  
The low-energy effective field theory analysis suggests that  spacetime supersymmetry is restored when expanding about the new vacuum.
However, the cancellation of anomalous effects between tree and one-loop level, and the breaking and restoration of supersymmetry means that these
models provide an interesting test case for our understanding of the subtleties of superstring perturbation theory. The current state of the art
is elucidated in papers by Witten \cite{Witten:2019uux} and by Sen \cite{Sen:2015uoa}.

Models with  pseudo-anomalous $\U(1)$ factors have been studied
in a variety of contexts in string theory beside the $\SO(32)$ theory with the standard embedding considered here. These include a broad class of orbifold compactifications \cite{Casas:1987us}, Type I and Type IIB string vacua \cite{Ibanez:1998qp}, M-theory \cite{March-Russell:1998lvm} and strongly coupled limits of the heterotic string \cite{Binetruy:1998hg}. There is also a vast literature on applications of these backgrounds to string model building and cosmology, for a small sample of the literature see \cite{Ibanez:1992fy, Ibanez:1994ig, Stewart:1994ts,Ramond:1995mb, Binetruy:1996xj, Irges:1998ax}. In this
paper will focus on the original context of the $\SO(32)$ heterotic string but expect that our analysis could be extended to this broader class of theories.

 In this work we extend the original analysis of the induced FI-like terms in \cite{Atick:1987gy,Dine:1987gj} to compute the one-loop renormalization of the
mass of perturbative BPS states.  We show that the mass corrections are non-zero
for states charged under the $\U(1)$ and, like the induced FI-like terms, the corrections are determined by the spectrum of massless states.  

This paper is organized as follows. In the second section we discuss massless and perturbative BPS states
for the $\SO(32)$ heterotic string on $\mcY \times S^1_R$  and discuss the vertex operators that we will use in this paper.  In the third section we review the computations in \cite{Atick:1987gy,Dine:1987gj} of one-loop corrections to the mass
of charged massless scalars and then extend them to compute the corrections to the masses of BPS states. We do this first using heuristic arguments, then
using a detailed orbifold computation, and finally we extend the super Riemann surface approach of \cite{Witten:2019uux} to our computations.  The final section contains our conclusions and a brief discussion of related work in progress.

\section{Massless and Massive BPS States and Their Vertex Operators}\label{sec:BPS}

In this paper we will use the covariant RNS formalism and the fermionic formulation of the $\SO(32)$ heterotic string. We also use the convention
where left-moving fields are anti-holomorphic and right-moving fields are holomorphic, and use the notation of denoting anti-holomorphic coordinates as $\tilde{z}$ rather than the standard $\bar{z}$. We start from a compactification on a Calabi-Yau threefold $\mcY$
with an embedding of the spin connection on $\mcY$ into the $\SO(32)$ gauge group. To describe this in the language of CFT, we start with the fermions
that describe the left-moving $\SO(32)$ affine Lie algebra in heterotic string theory. These fermions are in the $32$-dimensional vector representation
and under the embedding

\begin{equation}
\SO(32) \supset \SO(26) \times \SO(6) \supset \SO(26) \times \SU(3) \times \U(1)
\end{equation}
they decompose as

\begin{equation}
32 \rightarrow (26,1)_{0} + (1,3)_{-1} + (1, \bar 3)_{1}\label{eq:dc32}
\end{equation}
where the subscript refers to the $\U(1)$ charge. We denote the corresponding fermions by $\lambda^P$, $P=1, \dots 26$, $\lambda^i$ and $ \lambda^{\bar \imath}$
with $i, \bar \imath=1,2,3$.  When we identify the $\SU(3)$ gauge field with the $\SU(3)$-valued spin connection on $\mcY$ we end up with a $c=9$, $(2,2)$ supersymmetric
sigma-model ${\cal C}$ with coordinates $(\tilde Y^{i, \bar \imath}, Y^{i, \bar \imath}, \psi^{i, \bar \imath}, \lambda^{i, \bar \imath})$ where the $\psi^{i, \bar \imath}$ are the right-moving RNS fermion fields with components
tangent to $\mcY$. The full world-sheet CFT combines ${\cal C}$ with the remaining right and left-moving bosonic fields $X^\mu, \tilde X^\mu$, $\mu=0,1,2,3$, the right-moving
fermions $\psi^\mu$, and the remaining left-moving free fermions $\lambda^P$, as well as the standard complement of ghost fields for the heterotic string. It is customary to use the convention that the sigma-model fields on $\mcY$ be complex-valued, while all other fields are real-valued. This information is summarized in Table \ref{tab:orbcomp}.

\begin{table}[h]
	\centering
	\renewcommand{\arraystretch}{1.5}
	\begin{tabular}{| c | c |c|c|c|c|c|c|c|c|c|c|c|c|c|c|c|}
		\hline
		 & \multicolumn{4}{c|}{$\mu$\text{ (Spacetime)}} & \multicolumn{3}{c|}{$i,\bar{\imath}$\text{ (Calabi-Yau)}} & \multicolumn{6}{c|}{$i, \bar{\imath}, P$\text{ (Internal CFT)}} \\
		\hline
		& \multicolumn{4}{c|}{$\tilde X^{\mu}$, $\tilde b, \tilde c$} & \multicolumn{3}{c|}{$\tilde Y^{i, \bar \imath}$} & \multicolumn{3}{c|}{$\lambda^{i, \bar \imath}$} & \multicolumn{3}{c|}{$\lambda^{P}$} \\
		\hline
		\makecell{\text{L} \\ \text{(Bosonic String)}} & 0 & 1 & 2 & 3 & 1 &2 & 3 & 1 & 2 & 3 & 1 & \dots & 26 \\
		\hline
		\makecell{\text{R} \\ \text{(Superstring)}} & 0 & 1 & 2 & 3 & 1 & 2 & 3 &\multicolumn{6}{c|}{ }\\
		\hline
		& \multicolumn{4}{c|}{$ X^{\mu},{\psi}^{\mu}$, ${b}, {c}, {\beta},{\gamma}$} & \multicolumn{3}{c|}{$  Y^{i, \bar \imath}, {\psi}^{i, \bar \imath}$} &\multicolumn{6}{c|}{ } \\
		\hline
    	\end{tabular}
    	\caption{Sigma model fields and their indices in the Heterotic compactification on $\mcY$.}
    	\label{tab:orbcomp}
\end{table}

Although the computations that follow can be done without detailed knowledge of the CFT ${\cal C}$, it can be useful, as in the computations in \cite{Atick:1987gy}, to work in the orbifold limit of a particular Calabi-Yau threefold, namely the $Z$-manifold which can be treated using the techniques of orbifold conformal field theory \cite{Dixon:1985jw, Dixon:1986jc}.  In this construction one gauges a diagonal $\IZ_3$ subgroup of the $\IZ_3$ center of a $\SU(3)$ subgroup of $\SO(32)$ and a $\IZ_3$ subgroup of $\SO(6)$ acting on a six-torus $T^6= T^{2}_1 \times T^{2}_2 \times T^{2}_3$ where each two-torus $T^{2}_i$ has a $\IZ_3$ point group symmetry. This $\bbZ_{3}$ subgroup is generated by the matrix $\gamma = e^{2\pi i /3}\text{diag}(1,1,1) \in Z(\SU(3))$ and has a nontrivial action on the sigma model fields on $\mcY$. If we denote by $U$ any of the complex-valued fields $X^{i,\bar{\imath}}, \psi^{i,\bar{\imath}}, \tilde{X}^{i,\bar{\imath}}, \lambda^{i,\bar{\imath}}$, then the $\bbZ_{3}$ action is given by
\[U^{i} \mapsto \gamma^{ij} U^{j}\qquad U^{\bar\imath} \mapsto (\gamma^{*})^{\bar \imath \bar \jmath} U^{\bar\jmath}.\addtag\label{eq:twists}\]
Resolving the orbifold singularities leads to a Calabi-Yau threefold with Hodge numbers $h^{1,1}=36$ and $h^{2,1}=0$. If we work in the orbifold limit as we do here then the $\SU(3)$ subgroup of $\SO(32)$ remains unbroken and the gauge symmetry is $\SO(26) \times \SU(3) \times \U(1)$.

In either case, the string background preserves four-dimensional $\mcN=1$ supersymmetry and so there are no $0$-brane BPS states since
there are no central charges in the $\mcN=1$ supersymmetry algebra. There are however perturbative BPS $1$-branes or strings. A useful way to describe these states in string theory is to further compactify the theory to three dimensions on $S^1_R$, a circle of radius $R$. 
We write the left and right-moving momentum on $\IR^{2,1} \times S^1_R$ as $\tilde k=(k^\mu, \tilde k^3)$ and $k=(k^\mu, k^3)$, respectively, with $\mu=0,1,2$. Strings which wind
on the $S^1$ can be thought of, for very large $R$, as cosmic strings that wind around a large compact direction. 
 The left and right-moving string momenta on the $S^1_R$ are expressed in terms of the momentum $M$ and winding $L$ of the string as
\begin{equation}
\tilde k^3 = \frac{M}{R} + \frac{L R}{\alpha'}, \qquad k^3 = \frac{M}{R} - \frac{L R}{\alpha'} \, .\label{eq:k3lr}
\end{equation}
If we write $N_L, N_R$ for the sum of the usual left-moving and right-moving excitation
levels for the free fields, $(\tilde h, h)$ for the conformal dimensions of states in the CFT ${\cal C}$, and $k^\mu$ for the momenta in $\IR^{2,1}$, then the usual (three-dimensional) mass shell condition becomes
\begin{equation}
m^2= - k^\mu k_\mu = (\tilde k^3)^2 + \frac{4 (N_L -1)}{\alpha'}= (k^3)^2 + \frac{4}{\alpha'} \begin{cases} ( N_R -1/2) & ({\rm NS}) \\  N_R & ({\rm R}) \end{cases}\label{eq:3dmass}
\end{equation}

BPS states for which $m^2= (k^3)^2$ (and hence with $k^2=0$) are those with $ N_R=0$ in the Ramond sector and $ N_R=1/2$ in the Neveu-Schwarz sector.
Since these are the lowest-lying state that survive the GSO projection in the NS sector, it is simplest to just call these states right-moving ground states. It may not be completely obvious that these correspond to BPS states in spacetime. There are two ways to see that this is the case. First, spacetime supersymmetry charges are constructed from the right-moving degrees of freedom in the heterotic string. We will see that the vertex operators for these states have a right-moving part which is the same as that for massless states. As a result they fit into a supermultiplet of the same size as massless states, but since they are massive they must in fact be BPS. Second, one can construct the spacetime supersymmetry charges in string theory and compute their anticommutator and this yields a central term proportional to $k^{3}$. See for example section 11.6 of \cite{Polchinski:1998rr}. Thus the equation $m^2= (k^3)^2$ is the usual relation between the mass and central charge for BPS states.

For these BPS states we then have
\begin{equation}
m^2= \left( \frac{M}{R} - \frac{LR}{\alpha'} \right)^2= \frac{L^2 R^2}{\alpha'^2} + \frac{M^2}{R^2}- \frac{2ML}{\alpha'}, \qquad N_L -1 = -ML
\end{equation}
leading to an infinite tower of BPS states. For large $R$, the lightest states with nonzero winding are those with $N_L=0$ and $M=L=\pm 1$ and
\begin{equation}
m^2= \frac{R^2}{\alpha'^2}+\frac{1}{R^2} - \frac{2}{\alpha'}
\end{equation}
The one-loop renormalization of the mass of these BPS states was studied in \cite{Dabholkar:1989jt, Dabholkar:1990yf}, and, in agreement with those
papers, we will find that the one-loop renormalization of the mass of these states vanishes as a result of their being neutral under the pseudo-anomalous
$\U(1)$ gauge symmetry. 

The next lightest BPS states at large $R$ and with nonzero winding have $N_L=1$ and $M=0, L= \pm 1$ and have

\begin{equation}
m^2= \frac{R^2}{\alpha'^2} \, .
\end{equation}
These include states with nonzero $\U(1)$ charge and for these charged states we will see that their one-loop mass renormalization is nonzero.

Finally, while they will play no further role in this paper, there is an interesting class of extremal but not BPS states with $N_L=1$ which are
not in the right-moving ground state \cite{Dabholkar:1997fc}. Since they have $N_L=1$ their mass obeys an extremality condition relating the mass squared to charges,
\begin{equation}
m^2=  \left( \frac{M}{R} + \frac{LR}{\alpha'} \right)^2
\end{equation}
and the lightest such state with nonzero winding has mass squared
\begin{equation}
m^2= \frac{R^2}{\alpha'^2}+ \frac{1}{R^2} + \frac{2}{\alpha'} \, .
\end{equation}
The mass renormalization of these states was studied in \cite{Dabholkar:1997fc} and it would be interesting to extend their computations
to the backgrounds considered here with an anomalous $\U(1)$ gauge symmetry.

From now on we focus only on the BPS states and adopt the convention that $\alpha' = 2$.
Since these states are all in the right-moving ground state, the right-moving part of the vertex operators
creating these states are the same as the right-moving part of vertex operators creating massless states, the only change being that $\tilde k^3$ and $ k^3$ are
no longer the same when $ L \ne 0$. As a result, we will see that the computation of one-loop corrections to the masses of these states shares many similarities
to the computation of the corrections to massless states carried out in \cite{Atick:1987gy,Dine:1987gj}.

Since our goal is to compute the mass corrections for an infinite class of BPS states, we first make some general statements. In direct analogy with field theory, the mass renormalization of a general heterotic string state $\ket{V}$ at one loop arises as the first-order correction in $g_{s}$ to the on-shell propagator associated with $\ket{V}$. On the world-sheet, $\ket{V}$ is represented on the world-sheet by a weight $(1,1)$ superconformal vertex operator $V = V(\tilde k, k; \tilde z, z)$, so to obtain the mass renormalization of $\ket{V}$ we can simply evaluate the on-shell two-point function of $V$. Traditionally, this is accomplished by using vertex operators where the world-sheet supermoduli are integrated out at the outset. However, the on-shell two-point function is not strictly speaking well-defined when such vertex operators are used. There are two main workarounds to this issue. The first is to use the integrated vertex operators,  but to compute the two-point function with $V$ slightly off-shell. The answer is found to be constant in a neighborhood of the mass-shell condition, so we can simply define the on-shell result by that constant value. The justification for this procedure is that the off-shell propagator for $\ket{V}$ is well-defined in string field theory, and arises whenever $\ket{V}$ appears as an intermediate state of a superstring scattering amplitude. Thus, the answer to our question can be thought of as taking an on-shell limit. Interestingly, such a scattering amplitude is nonzero only when $\ket{V}$ is very nearly on-shell. In the language of analytic geometry, the amplitude receives a nonzero contribution only at the boundary of the full (Deligne-Mumford compactified) moduli space of Riemann surfaces. To fully establish the validity of this approach, one needs to find a set of external on-shell states for which the scattering amplitude has $V$ as an isolated intermediate channel. In \cite{Atick:1987gy}, this was explicitly done for the massless state with $N_{L}=1$, but there is no known procedure to accomplish this for general $\ket{V}$. Nevertheless, the strategy of computing correlators slightly off-shell is conceptually more straightforward and is what we will use in the coming sections. The other approach, which benefits from the fact that we can take $V$ to be on-shell, instead analyzes the problem in a manifestly supersymmetric way by using super Riemann surfaces \cite{Witten:2012bh, Witten:2012ga}. This treatment is natural from the standpoint of the heterotic world-sheet, which has $(0,1)$ supersymmetry. Using this formalism, one arrives at the same answer after evaluating in super-moduli space the superspace analogue of a conditionally-convergent integral \cite{Witten:2019uux}. We will comment on this formalism in section \ref{ssec:susyform} and unsurprisingly, find agreement between these two approaches.

We now consider the vertex operators for the states whose mass renormalization we want to study.  For convenience we will often work in the $\IZ_3$ orbifold
although our computations can be easily extended to more general Calabi-Yau threefolds. 
In the $\IZ_3$ orbifold 
there are massless scalar states with $M=L=0$ and $N_L=1$ in the representation $(26,  \bar 3)_{1}$ which arise in the NS sector for fermions on both
the left and right which have the form
\begin{equation}
\lambda^P_{-1/2} \lambda^{\bar{\jmath}}_{-1/2} |\tilde k \rangle_L \otimes \psi^{j}_{-1/2} |  k \rangle_R
\end{equation}
with the corresponding vertex operator in the $-1$ picture given by
\begin{equation}
V^{P}(\tilde k, k; \tilde{z}, z) = i \lambda^P(\tilde z) \lambda^{\bar{\jmath}}(\tilde z) e^{i\tilde k \cdot \tilde X} e^{-\phi}(z) \psi^{j}(z)e^{ik\cdot X}  \,  \label{eq:mlscalar}
\end{equation}
In the $0$ picture this vertex operator becomes
\begin{equation}
V^P(\tilde k, k; \tilde{z}, z) = i \lambda^P(\tilde z) \lambda^{\bar{\jmath}}(\tilde z) e^{i\tilde k \cdot \tilde X} \left( \partial X^{j}(z) + i(k\cdot \psi)(z) \psi^{j}(z) \right) e^{ik\cdot X}
\end{equation}
and the vertex operator for the conjugate field $\bar{V}$ in the $(26, 3)_{-1}$ representation is given by sending $\tilde k, k\mapsto -\tilde k, -k$ and exchanging barred and unbarred $\SU(3)$ indices,
\begin{equation}
\bar{V}^P(-\tilde k, -k; \tilde{z}, z) = -i \lambda^P(\tilde z) \lambda^{j}(\tilde z) e^{- i\tilde k \cdot \tilde X}\left(\partial X^{\bar{\jmath}}(z) - i(k\cdot \psi)(z) \psi^{\bar{\jmath}}(z) \right) e^{-ik\cdot X}
\end{equation}
In these vertex operators the left and right-moving momenta ($\tilde k, k)$ are equal since $L=0$.
In what follows, the conventions and the form of the vertex operators agree with \cite{Atick:1987gy}, except that we keep track of the separate dependence on $\tilde k$ and $k$.

In this paper we mainly focus on the mass correction to massive BPS states. The right-moving part of the vertex operator for these states, which we denote by $V_{R}$, is the same as the original massless state
except that since now $L \ne 0$ the left and right-moving momenta $\tilde k, k$ will be unequal due to \eqref{eq:k3lr}.
The left-moving part of the vertex operator we write as
\begin{equation}
V_L^{P\bar{\jmath}}(\tilde k; \tilde z) = O^{(\tilde{h},Q)P\bar{\jmath}}(\tilde z)e^{i\tilde k \cdot \tilde X}
\end{equation}
where $O^{(\tilde{h},Q)P\bar{\jmath}}$ is a primary operator for the left-moving Virasoro algebra with conformal weight $\tilde{h}$ and charge $Q$ under the left-moving $\U(1)$ current $J_{\ell}$, given by $J_\ell= \lambda_i \lambda^i$ in the orbifold limit. We let $\bar O^{(\tilde{h},-Q)P j}$ denote the vertex operator for the conjugate states of charge $-Q$.
Thus we have
\begin{equation}
O^{(\tilde{h},Q)P\bar{\jmath}}(\tilde z_{1}) \bar O^{(\tilde{h},-Q)P'j}(\tilde z_{2})  \sim \frac{\delta^{PP'}\delta^{\bar{\jmath}j}}{(\tilde z_{1} -\tilde z_{2})^{2 \tilde{h}}} + \cdots
\end{equation}
and
\begin{equation}
J_\ell(\tilde z_{1}) O^{(\tilde{h},Q)P\bar{\jmath}} (\tilde z_{2}) \sim \frac{Q O^{(\tilde{h},Q)P\bar{\jmath}}(\tilde z_{2})}{\tilde z_{1} -\tilde z_{2}}\label{eq:JOope}
\end{equation}
For most of this paper we will work directly with the operators $O^{(\tilde h, Q) P\bar \imath}$, but in some cases it is useful to factor the operator $O^{(\tilde h, Q) P\bar \imath} = \normord{O^{(\tilde h - 1/2, Q-1)P}_{\text{aux}} \lambda^{\bar \imath}}$. We will also often omit $\SO(26)$ indices of $O^{(\tilde{h},Q)P\bar{\jmath}} =: O^{(\tilde{h},Q)\bar{\jmath}}$, $O^{(\tilde h - 1/2, Q-1)P}_{\text{aux}} =: O^{(\tilde h - 1/2, Q-1)}_{\text{aux}}$, and $V^{P}=: V$ but it should be understood that in general these operators transform in some nontrivial representation of $\SO(26)$.

Before we conclude this section, let us mention a relevant feature of superstring perturbation theory. The conformal dimension of a world-sheet vertex operator and the spacetime rest mass of the string state which it creates are related by conformal invariance. A nonzero mass renormalization is therefore in conflict with the usual formulation of string perturbation theory built on conformal invariance since any deviation $\delta m^{2}$ in mass explicitly breaks conformal symmetry. One resolution, as explained in \cite{Seiberg:1986ea}, is to dress the operator $V$ order-by-order in $g_{s}$ with an appropriate wave-function renormalization proportional to $\delta m^{2}$ at that order in $g_{s}$. This cancels the deviation in weight due to the renormalized mass and restores conformal symmetry at the quantum level. Alternatively, this issue can be dealt with using string field theory \cite{Sen:2015uoa}.

\section{One-loop Mass Corrections}\label{sec:oneloop}

In this section, we present our computations for the one-loop mass correction to BPS states with nonzero $\U(1)$ charge. 
For simplicity we often consider the heterotic string compactified on an $\bbZ_{3}$-orbifold, but the results should generalize to arbitrary Calabi-Yau threefolds by similar arguments to those given in \cite{Atick:1987gy, Dine:1987gj, Witten:2019uux}. We start with a heuristic explanation of the mass correction which generalizes the arguments
given in \cite{Atick:1987gy, Dine:1987gj}. We then turn to a detailed orbifold computation that generalizes the result of \cite{Atick:1987gy}. These computations
require working slightly off-shell, but in this formulation we can identify the left-moving coordinate $\tilde{z}$ with the complex conjugate $\bar{z}$ of the right-moving one $z$. We then remedy this by generalizing the computation of \cite{Witten:2019uux} using super-Riemann surfaces to obtain the same result, but from a purely on-shell computation. 

\subsection{Heuristics}

We start with a brief summary of the one-loop mass correction to the massless charged scalars created by the vertex operator $V^{P}(\tilde{k}, k; z) := V^{P}(\tilde k, k; \tilde{z} = \bar{z}, z)$. After using translation
invariance on the torus $\Sigma$ to set the location of one vertex operator to $z'=0$, the mass correction arises from the two-point function
\begin{equation}
\int_{\Sigma} d^2 z \langle V^{P}(\tilde k, k; z) \bar{V}^{P'}(-\tilde k, - k; 0) \rangle
\end{equation}
after summing over spin structures on $\Sigma$ and integrating over the modular parameter $\tau$ of the torus.
In the product of the two vertex operators there is a term with no $\psi$ fields, terms with two $\psi$ fields and a term with four $\psi$ fields.
As explained in  \cite{Atick:1987gy, Dine:1987gj}, the term with no $\psi$'s factorizes into a bosonic part times the fermion partition function and the latter vanishes after summing over spin structures as a result of spacetime supersymmetry. The terms with two $\psi$'s are linear in $\psi^\mu$ and so vanish as a result of four-dimensional Lorentz invariance.
The remaining term with four fermions involve the correlation function of fermions with four-dimensional spacetime indices
$\langle (k\cdot \psi)(k\cdot \psi) \rangle $, which is proportional to $k^2$, and thus appears to vanish on-shell for massless states. However, this vanishing is an illusion because the integral over $z$ has a $1/k^2$ singularity arising from a term appearing in  the OPE of the two vertex operators as $z \rightarrow 0$. This makes the integral ambiguous. One way to deal with this is to work slightly off-shell, with $k^2$ nonzero, and then take the limit $k^2 \rightarrow 0$ at the end of the computation. 

The factor of $1/k^2$ arises as follows. It was explained in  \cite{Atick:1987gy} that there is an operator $V_D$ which can be thought of as the vertex operator for the
auxiliary field $D$ associated to the pseudo-anomalous $\U(1)$ gauge symmetry. In the orbifold limit, this vertex operator takes the simple form $V_D= J_\ell J_r$ where

\begin{equation}
J_\ell = \lambda_i \lambda^i, \qquad J_r = \psi_i \psi^i \, .
\end{equation}
In the correlation function $ \langle V^P(\tilde k, k; z) \bar{V}^{P'}(-\tilde k, -k; 0)\rangle $ we use the OPE's

\begin{equation}
e^{i (\tilde k \cdot \tilde X + k \cdot X)}(z) e^{- i(\tilde k\cdot \tilde X + k \cdot X)}(0) \sim \frac{1}{ |z|^{2k^2}} \, ,
\end{equation}
\begin{equation}
\lambda^P(\tilde z) \lambda^{P'}(0) \sim \frac{\delta^{PP'} }{\tilde z} \, ,
\end{equation}
and

\begin{equation}
(k \cdot \psi)(z) (k \cdot \psi)(0) \sim \frac{k^2}{z} \, ,
\end{equation}
which yields
\begin{equation}
\int d^2z \frac{k^2}{|z|^{2+2k^2}} \langle V_D \rangle \, .
\end{equation}
The integral over $z$  diverges for small $k^2$ as $1/k^2$ and when combined with the factor of $k^2$ gives, in the limit that $k^2 \rightarrow 0$ an answer
proportional to $\langle V_D \rangle$. This latter quantity can be evaluated using the operator formalism, and as was shown in \cite{Atick:1987gy, Dine:1987gj},
it depends only on the massless spectrum of the theory and is proportional to the sum of the $\U(1)$ charges of the massless multiplets weighted by their
helicity and is nonzero whenever $\chi(\mcY) \ne 0$. The exact coefficient of the one-loop mass renormalization requires a more detailed computation and was
first done in \cite{Atick:1987gy}. We will generalize this computation to our case in the following subsection. 

This heuristic explanation for the non-zero mass renormalization for massless states can be extended to the  perturbative BPS states described in
the preceding section provided that we are careful to distinguish the left and right-moving momenta and that we take account the effects 
of the conformal dimension of the left-moving part of the vertex operators when computing the OPE.

The vertex operators for BPS states  have the form
\begin{equation}
V(\tilde k, k; z):= V^{(\tilde{h},Q)}(\tilde k, k; z) = O^{(\tilde{h},Q)\bar{\jmath}} e^{i\tilde k \cdot \tilde X} \left(\partial X^{j}(z) + i (k \cdot \psi)(z) \psi^{j}(z) \right) e^{i k \cdot X} \,.
\end{equation}
The operator $O^{(\tilde{h},Q)}$ has conformal dimension $(\tilde{h},0)$ and $\U(1)$ charge $Q$ and we have been careful to distinguish left and right-moving momenta $\tilde k, k$. To survive the orbifold projection it is important that $O^{(\tilde{h},Q)} = O^{(\tilde{h},Q)\bar{\jmath}}$ transform with phase $e^{-2 \pi i/3}$ under the center of $\SU(3)$ or equivalently, have charge $Q \equiv 1~\mod~3$; in the following section we explicitly construct states which satisfy this condition. $O^{(\tilde{h},Q)}$ may also carry $\SO(26)$ quantum numbers, but they play no distinguished role in the analysis. For simplicity of exposition we will work with $\SO(26)$ singlets and take $O^{(\tilde h, Q)}$ to be a primary operator in the sigma-model $\mcC$, but it is not hard to generalize our computations to general $O^{(\tilde h, Q)}$.
As before, the two-point function
\begin{equation}
\int_{\Sigma} d^2 z \langle V^{(\tilde{h},Q)}(\tilde k, k; z) \bar{V}^{(\tilde{h},-Q)}(-\tilde k, -k; 0) \rangle \, ,
\end{equation}
after summing over spin structures only receives contributions from the term involving four $\psi$ fields. Again, this term naively seems
to vanish because $\langle (k\cdot\psi) (k\cdot \psi) \rangle  \propto k^2$ and the on-shell relation $k^2 = k^\mu k_\mu + (k^{3})^2=-m^2 + (k^{3})^2=0$ for BPS states. However, we can again
use OPE's to see that there is a compensating factor of $1/ k^2$ from the integral over $z$ when we compute the coefficient of $V_D$ in the OPE. We now have
\begin{equation}
e^{i \tilde k \cdot \tilde X(\tilde z)} e^{-i \tilde k \cdot \tilde X(0)} \sim \frac{1}{ {\tilde z}^{\tilde k^2}} \, ,
\end{equation}
\begin{equation}
e^{i k \cdot X(z)} e^{-ik \cdot X(0)} \sim \frac{1}{ z^{k^2}} \, ,
\end{equation}
\begin{equation}
O_{\text{aux}}^{(\tilde{h}-1/2,Q-1)}(\tilde z) \bar O_{\text{aux}}^{(\tilde{h}-1/2,1-Q)}(0) \sim \frac{1}{\tilde z^{2\tilde{h}-1}} \, ,
\end{equation}
where $O_{\text{aux}}$ was defined under equation \eqref{eq:JOope}, and
\begin{equation}
(k \cdot \psi)(z) (k \cdot \psi)(0) \sim \frac{ k^2}{ z} \, ,
\end{equation}
which yields
\begin{equation}
\int d^2z \frac{k^2}{\tilde z^{\tilde k^2+2\tilde{h}-1}  {z}^{1+ k^2}} \langle V_D \rangle \, .
\end{equation}
Assuming that we can go slightly off-shell while maintaining the level-matching condition $\tilde k^2+ 2\tilde{h}-2= k^2$, this becomes
\begin{equation}
\int d^2 z \frac{k^2}{ |z|^{2+ 2 k^2}} \langle V_D \rangle
\end{equation}
and again we get a result proportional to $\langle V_D\rangle $ as $k^2 \rightarrow 0$.

To compute the exact coefficient of the mass renormalization of these BPS states requires a more detailed computation which we choose to
do using an orbifold construction following the analysis in the second part of \cite{Atick:1987gy}. This is the subject of section \ref{eq:orbcomp}. 
The requirement of working slightly off-shell can be avoided by using the formalism of super Riemann surfaces. We give this analysis following
our discussion of the orbifold computation.

\subsection{Orbifold Computation of BPS Mass Correction}\label{eq:orbcomp}

Our starting point for the orbifold computation of the mass renormalization of BPS states is the gauge-fixed two-point function 
\[m^{2}_{\text{one-loop}} = - g_{s}^{2} \int \frac{\dd[2]{\tau}}{4\Im \tau} \int \dd[2]{z_{1}}\dd[2]{z_{2}} \left\langle \tilde c(0)c(0)b(0)\tilde b(0) V(\tilde k, k; z_{1})\bar{V}(-\tilde k, -k; z_{2}) \right\rangle_{T^{2}}.\addtag\label{eq:mlcorr}\]
Since we have integrated superspace vertex operators over odd moduli, it will be easiest to work with the 0 picture vertex operators, which take the form

\[V(\tilde k, k; z) = O^{(\tilde{h}, Q)\bar{\jmath}}(\tilde k; \tilde z) e^{i\tilde k\cdot \tilde X} \qty(\partial X^{j}(z) + i(k\cdot \psi) ({z})\psi^{j}(z)) e^{i k\cdot X}.\addtag\label{eq:genV}\]
In equation \eqref{eq:genV}, $O^{(\tilde{h}, Q)}$ is a primary operator whose holomorphic weight $\tilde{h}$ satisfies $\tilde k^{2} + 2\tilde h -2 = k^{2}$ at tree level. For convenience we denote
\[V^{\bar{\jmath}}_{L}(\tilde k;\tilde z) = O^{(\tilde{h},Q)\bar{\jmath}} (\tilde k;\tilde z) e^{i\tilde k\cdot \tilde X},\qquad V^{j}_{R}(k; z) = \qty(\partial X^{j}(z) + i(k\cdot \psi)(z)\psi^{j}(z)) e^{i k\cdot X}.\addtag\]

An important step when computing orbifold correlation functions is to project the Hilbert space onto the subspace of states invariant under the discrete group action. Due to modular invariance, one must also include twisted sector states, which are themselves invariant under the group action but are allowed to have nontrivial holonomy valued in the discrete group. This procedure is essentially that of gauging a target space discrete symmetry. For the $Z$-manifold, it is a discrete $\bbZ_{3}$-symmetry that is being gauged, whose action on the sigma model fields was described in the previous section. The full Hilbert space of states can therefore be organized according to their twists $(e^{2\pi i c_{j}}, e^{2\pi i d_{j}})$ in the target torus $T^{2}_{j}$, $j=1,2,3$, where $c_{j}, d_{j} \in \frac{1}{3}\bbZ/\bbZ$ are the twists when going around the space-like and time-like cycles on the world-sheet, respectively. Fixing a space-like twist $c_{i}$ on $T^{2}_{i}$, the projection onto $\bbZ_{3}$-invariant states can be carried out using the operator $\frac{1}{3}(1 + g + g^{2})$, where $g\in \bbZ_{3}$ acts on the fields by a $e^{2\pi i d_{j}}$ twist as in \eqref{eq:twists}. For the twists $(\mathbf{c}, \mathbf{d})$ on $T^{6}$ to embed in the center of $\SU(3)$, and hence for the orbifold compactification to preserve $\mcN=1$ supersymmetry in four dimensions, the twists should satisfy the constraints $ \sum_{i}c_{i} = \sum_{i}d_{i} \equiv 0~\mod~ \bbZ$ \cite{Dixon:1986qv, Dixon:1985jw, Dixon:1986jc}. This leads to three sets of space-like boundary conditions $(c_{1}, c_{2}, c_{3}) = (\frac{1}{3}n, \frac{1}{3}n, \frac{1}{3}n)$. States with $n=0$ lie in the untwisted sector, while those with $n=1$ and $n=2$ are in twisted sectors. If we denote a twist structure on the $Z$-manifold by ${\mathbf{c}\brack \mathbf{d}}$ and the GSO-symmetric correlator evaluated in that sector by $ \langle V\tilde{V} \rangle_{T^{2}}^{\qty(\mathbf{c}\brack \mathbf{d})}$, then the full correlator before integrating over moduli is given by
\[ \left\langle V\tilde{V} \right\rangle_{T^{2}} = \frac{1}{3} \sum_{(\mathbf{c}, \mathbf{d})} \left\langle V\tilde{V} \right\rangle_{T^{2}}^{\qty({\mathbf{c}\brack \mathbf{d}})}.\addtag\label{eq:orbsum}\]

Substituting standard expressions from the literature \cite{Alvarez-Gaume:1986rcs, DHoker:1988pdl, Polchinski:1998rr}, it is not difficult to show that many terms in the correlator \eqref{eq:mlcorr} vanish by theta function identities after summing over right-moving spin structures. The cancellation is very similar to that which results in the vanishing of the one-loop superstring cosmological constant in ten dimensions. The only term that survives is $ \int \dd[2]{z} \left\langle (k\cdot \psi)\psi^{j}(z) ( k\cdot \psi) \psi^{\bar{\jmath}} (0) \right\rangle$ and is proportional to $k^{2}$, which vanishes in the on-shell limit. Despite this, the full correlator \eqref{eq:mlcorr} need not be zero \cite{Atick:1987gy, Dine:1987gj}. The heuristic explanation of this fact was given in the previous section. A more geometric explanation is that in superstring perturbation theory one integrates over the full (Deligne-Mumford compactified) moduli space of Riemann surfaces, which contains infrared singularities. If the integrand has a pole with residue $1/k^{2}$, representing an infrared effect where the intermediate state $\ket{V}$ goes on-shell \cite{Friedan:1986ua,Witten:2019uux}, then the contribution from this point is capable of cancelling the factor of $k^{2}$ above to give a finite nonzero answer. Therefore, one of our goals is to extract the contributions from these poles, which we interchangeably describe as ``the on-shell limit'' or ``the limit of coincident points.''

In order to evaluate the one-loop correlator, it is instructive to work out the result for the left and right movers separately, then combine them at the end of the computation. We begin with the right-moving/holomorphic part. Projecting onto GSO-invariant states and summing over GSO-twisted sectors, the holomorphic correlator with orbifold twist structure $(\mathbf{c}, \mathbf{d})$ takes the form
\begin{multline}
\left\langle V^{j}_{R}(z_{1})\bar{V}^{\bar{\jmath}}_{R}(z_{2}) \right\rangle_{T^{2}}^{\qty({\mathbf{c}\brack \mathbf{d}})} = \frac{ k^{2}\delta^{j\bar{\jmath}}}{16\pi^{2}\Im \tau}  \frac{e^{ k^{2}g({z}_{1}, {z}_{2})}}{\eta(\tau)^{3}} \frac{\theta_{1}'(0)^{2}}{\theta_{1}^{2}(z_{1} - z_{2})}\qty(\theta{1/2 + c_{1}\brack 1/2 + d_{1}}(0)\theta{1/2 + c_{2}\brack 1/2 + d_{2}}(0)\theta{1/2 + c_{3}\brack 1/2 + d_{3}}(0))^{-1}\\
\times \sum_{(a, b)} \delta_{(a, b)}\theta{a \brack b}(z_{1} - z_{2}) \theta{a + c_{1}\brack b + d_{1}}(z_{1}- z_{2}) \theta{a + c_{2}\brack b + d_{2}}(0) \theta{a + c_{3}\brack b + d_{3}}(0)\label{eq:rmcorr}
\end{multline}
where $\delta_{(a, b)}$ is a spin structure-dependent phase equal to 1 and -1 for even and odd spin structures, respectively. In the untwisted sector the above expression is supplemented with an additional factor of $\prod_{j=1}^{3} (1 - e^{2\pi i d_{j}})$ after a careful treatment of bosonic zero modes on the orbifold. As a result the correlator with trivial twist $(\mathbf{c}, \mathbf{d}) = (\mathbf{0}, \mathbf{0})$ is zero. The expression $g$ in equation \eqref{eq:rmcorr} is the propagator of the chiral boson which transforms non-trivially under the mapping class group $\SL(2,\bbZ)$. If we denote the Jacobian of the torus by $J(T^{2})$, $g$ is the unique section, up to rescaling, of a holomorphic line bundle over $J(T^{2})\times J(T^{2})$ pulled back along the Abel-Jacobi map $T^{2}\to J(T^{2})$. For the purpose of calculating superstring amplitudes, we can approximate $g$ (and its anti-holomorphic counterpart $\tilde{g}$) by the prime form $E$,

\[g(z_{1}, z_{2}) = -\log E(z_{1}, z_{2}) = -\log\frac{\theta_{1}(z_{1}-z_{2})}{\theta'_{1}(0)},\qquad \tilde{g}(\tilde{z}_{1}, \tilde{z}_{2}) = g(z_{1}, z_{2})^{*}\addtag\]
which captures the pertinent information of the correlator in the on-shell limit. Using the generalized Riemann identity \cite{Mumford:1983}, the sum over spin structure in equation \eqref{eq:rmcorr} can be performed explicitly. The final result takes the form

\[\left\langle V^{j}_{R}(z_{1})\bar{V}^{\bar{\jmath}}_{R}(z_{2}) \right\rangle_{T^{2}}^{\qty({\mathbf{c}\brack \mathbf{d}})} = \frac{k^{2}\delta^{j\bar{\jmath}}}{8\pi^{2}\Im \tau}  \frac{e^{k^{2}g({z}_{1}, {z}_{2})}}{\eta(\tau)^{3}} \frac{\theta_{1}'(0)^{2}}{\theta{1/2 + c_{1}\brack 1/2 + d_{1}}(0)} \frac{\theta{1/2 + c_{1}\brack 1/2 + d_{1}}(z_{1}-z_{2})}{\theta_{1}(z_{1} - z_{2})}.\addtag\]

We now move onto the left-moving/anti-holomorphic fields. It is possible to say something general about the correlator here even though the exact form of the vertex operator will depend on the BPS state we are studying. Fixing a twist $(\mathbf{c}, \mathbf{d})$ and left-moving spin $(a',b')$ structure, consider the Laurent expansion of the following correlator in the limit of coincident points

\[ \left\langle O^{(\tilde{h}, Q)}(\tilde z_{1})\bar{O}^{(\tilde{h}, -Q)}(\tilde z_{2}) \right\rangle_{T^{2}}^{\qty({a'\brack b'}, {\mathbf{c}\brack \mathbf{d}})} = (\tilde z_{1}-\tilde z_{2})^{-2\tilde{h}} \sum_{n \geq 0} F_{n}^{\qty({a'\brack b'},{\mathbf{c}\brack \mathbf{d}})}(\tilde z_{1}-\tilde z_{2})^{n}.\label{eq:laur}\addtag\]
The coefficients $F_{n}$ are c-numbers which depend on twist and left-moving spin structure. The leading coefficient $F_{0}$ is captured entirely by the OPE, while subleading terms in the expansion correct for the topology of the torus. Using standard expressions, we can then write down the left-moving part of the correlator in a given orbifold twist structure $(\mathbf{c}, \mathbf{d})$:

\begin{multline}
\left\langle V^{\bar{\jmath}}_{L}(\tilde{z}_{1})\bar{V}^{j}_{L}(\tilde{z}_{2}) \right\rangle_{T^{2}}^{\qty(\mathbf{c}\brack \mathbf{d})} = \frac{\delta^{\bar{\jmath}j}}{16\pi^{2}\Im\tau} \qty{\frac{e^{\tilde k^{2}g(z_{1}, z_{2})}}{\eta(\tau)^{15}} \qty(\theta{{1/2 + c_{1}}\brack {1/2 + d_{1}}}(0)\theta{{1/2 + c_{2}}\brack {1/2 + d_{2}}}(0)\theta{{1/2 + c_{3}}\brack {1/2 + d_{3}}}(0))^{-1}}^{*}\\
\times (\tilde z_{1}-\tilde z_{2})^{-2\tilde{h}} \sum_{n\geq 0} (\tilde z_{1}-\tilde z_{2})^{n} \sum_{(a',b')} F_{n}^{\qty({a'\brack b'},{\mathbf{c}\brack \mathbf{d}})} \qty{\theta{a' + c_{1}\brack b' + d_{1}}(0)\theta{a' + c_{2}\brack b' + d_{2}}(0)\theta{a' + c_{3}\brack b' + d_{3}}(0)\theta^{13}{a'\brack b'}(0)}^{*}
\end{multline}
This expression cannot be simplified further without additional details on the vertex operators. Just as before, there is an additional factor of $\prod_{j=1}^{3} (1 - e^{-2\pi i d_{j}})$ in the untwisted sector. In the full correlator \eqref{eq:mlcorr} these phases contribute an overall factor of $\prod_{j=1}^{3} \abs{1 - e^{2\pi i d_{j}}}^{2} = 27$ in the untwisted sector, while for each of the twisted sectors the same factor arises after summing over the Hilbert spaces at the $3^{3}=27$ fixed points on the $Z$-manifold. After projecting onto $\bbZ_{3}$-invariant states, including twisted sector states, and integrating over moduli, we find that
\[\begin{aligned}
m^{2}_{\text{one-loop}} =& \frac{9g_{s}^{2}k^{2}}{128\pi^{4}} \int \frac{\dd[2]{\tau}}{4(\Im \tau)^{3}} \frac{\theta_{1}'^{2}(0)}{[\eta(\tau)^{15}]^{*}\eta(\tau)^{3}} \sum_{n\geq 0} \int \dd[2]{z_{1}} \dd[2]{z_{2}} (\tilde z_{1}-\tilde z_{2})^{-2\tilde{h} + n}\frac{e^{\tilde k^{2}\tilde g(\tilde z_{1}, \tilde z_{2})}e^{ k^{2}{g}({z}_{1}, {z}_{2})}}{\theta_{1}(z_{1}-z_{2})}\\
&\qquad \times \sum_{(\mathbf{c}, \mathbf{d})} \frac{\theta{1/2 + c_{1}\brack 1/2 + d_{1}}(z_{1}-z_{2})}{\theta{1/2 + c_{1}\brack 1/2 + d_{1}}(0)}\qty{\qty( \theta{1/2 + c_{1}\brack 1/2 + d_{1}}(0)\theta{1/2 + c_{2}\brack 1/2 + d_{2}}(0)\theta{1/2 + c_{3}\brack 1/2 + d_{3}}(0))^{-1}}^{*} \\
&\qquad \times \sum_{(a',b')} F_{n}^{\qty({a'\brack b'},{\mathbf{c}\brack \mathbf{d}})} \qty{\theta{a' + c_{1}\brack b' + d_{1}}(0)\theta{a' + c_{2}\brack b' + d_{2}}(0)\theta{a' + c_{3}\brack b' + d_{3}}(0)\theta^{13}{a'\brack b'}(0)}^{*}
\end{aligned}\addtag\label{eq:totcorr}\]

Although the order of the singularity in $\tilde z_{1}-\tilde z_{2}$ arising from the correlators of left-moving bosons and fermions separately depend on $\tilde{h}$, the tree-level constraint $\tilde k^{2} + 2\tilde h - 2 = 0$ ensures that the combination of $\tilde{z}_{1} - \tilde{z}_{2}$ appearing in equation \eqref{eq:totcorr} ends up having no $\tilde{h}$-dependence. To extract the desired factor of $1/k^{2}$, the following identity for a conditionally-convergent integral \cite{Witten:2012bh, Witten:2019uux} evaluated as $k^{2}\to 0$ is useful:

\[\int \dd[2]{z_{1}}\dd[2]{z_{2}} \frac{1}{(\tilde z_{1}-\tilde z_{2})^{s + k^{2}}(z_{1}-z_{2})^{1 + k^{2}}} = \frac{\pi}{k^{2}} \delta_{s,1}.\addtag\]
The delta function arises from performing the angular part of the integral first, and in equation \eqref{eq:totcorr} it picks out from the infinite sum of terms the one with $n = 1$. It follows that the one-loop mass correction is equal to

\[\begin{aligned}
m^{2}_{\text{one-loop}} &= - \frac{9g_{s}^{2}}{64\pi^{2}} \int \frac{\dd[2]{\tau}}{(\Im\tau)^{2}} \frac{1}{[\eta(\tau)^{15}]^{*}} \sum_{(\mathbf{c}, \mathbf{d})} \qty{\qty( \theta{1/2 + c_{1}\brack 1/2 + d_{1}}(0)\theta{1/2 + c_{2}\brack 1/2 + d_{2}}(0)\theta{1/2 + c_{3}\brack 1/2 + d_{3}}(0))^{-1}}^{*} \\
&\qquad \times \sum_{(a',b')} F_{1}^{\qty({a'\brack b'},{\mathbf{c}\brack \mathbf{d}})} \qty{\theta{a' + c_{1}\brack b' + d_{1}}(0)\theta{a' + c_{2}\brack b' + d_{2}}(0)\theta{a' + c_{3}\brack b' + d_{3}}(0)\theta^{13}{a'\brack b'}(0)}^{*}
\end{aligned}\label{eq:oneloop}\addtag\]
where we have used the relation $\theta'_{1}(0) = -2\pi \eta(\tau)^{3}$.

We will use this general result to evaluate the one-loop mass renormalization of massive BPS states in the next section, but first as a simple check, we use it to compute the mass renormalization for the massless state studied in \cite{Atick:1987gy}. The RNS zero-picture vertex operator associated with this state is given by

\[V(\tilde k, k; z) = i\lambda^{P}(\tilde{z})\lambda^{\bar{\jmath}}(\tilde{z}) e^{i\tilde k\cdot \tilde X} \qty(\partial X^{j}(z) + i(k\cdot \psi)(z)\psi^{j}(z)) e^{i k\cdot X}\addtag\]
where $P$ is an $\SO(26)$ index. In this case, the operator $O^{(\tilde{h}, Q)P\bar{\jmath}}$ is simply the left-moving fermion current $J_{L}^{P\bar{\jmath}} = i\lambda^{P}\lambda^{\bar\jmath}$ and has weight $(\tilde h, h) = (1,0)$, so we have $\tilde k^{2} = k^{2} - 2\tilde{h} + 2 = 0$ in the on-shell limit. Substituting standard expressions for fermionic propagators, one finds

\[ \left\langle O^{(\tilde{h}, Q)P\bar{\jmath}}(\tilde z_{1})\bar{O}^{(\tilde{h}, -Q)P'j}(\tilde z_{2}) \right\rangle^{\qty({a'\brack b'}, {\mathbf{c}\brack \mathbf{d}})}_{T^{2}} = \delta^{\bar{\jmath}j}\delta^{PP'}\qty{\frac{\theta_{1}'^{2}(0)\theta{a' \brack b'}(z_{1}-z_{2}) \theta{a' + c_{1}\brack b' + d_{1}}(z_{1}-z_{2})}{\theta{a' \brack b'}(0)\theta{a' + c_{1}\brack b'+d_{1}}(0)\theta^{2}_{1}(z_{1}-z_{2})}}^{*}\addtag\]
and a short computation shows that the coefficient $F_{1}^{\qty({a'\brack b'}, {\mathbf{c}\brack \mathbf{d}})}$ that enters into equation \eqref{eq:oneloop} is

\[F_{1}^{\qty({a'\brack b'}, {\mathbf{c}\brack \mathbf{d}})}(\tau) = \qty{\frac{\theta'{a' + c_{1}\brack b' + d_{1}}(0)}{\theta{a' + c_{1}\brack b' + d_{1}}(0)} + \frac{\theta'{a' \brack b'}(0)}{\theta{a' \brack b'}(0)}}^{*}.\addtag\]
It follows that the one-loop mass correction to the massless state is equal to

\[m^{2}_{\text{one-loop}} = -\frac{9g_{s}^{2}}{64\pi^{2}} \int \frac{\dd[2]{\tau}}{(\Im\tau)^{2}} \qty(\frac{1}{\eta(\tau)^{15}} \sum_{(\mathbf{c}, \mathbf{d})} \sum_{(a',b')}\frac{\theta'{a' + c_{1}\brack b' + d_{1}}(0)\theta{a' + c_{2}\brack b' + d_{2}}(0)\theta{a' + c_{3}\brack b' + d_{3}}(0)\theta^{13}{a'\brack b'}(0)}{\theta{1/2 + c_{1}\brack 1/2 + d_{1}}(0)\theta{1/2 + c_{2}\brack 1/2 + d_{2}}(0)\theta{1/2 + c_{3}\brack 1/2 + d_{3}}(0)})^{*} = 12g_{s}^{2}. \addtag\label{eq:inttau}\]
In the second equality, we have used the fact that the quantity in the parentheses is a bounded modular form of weight zero and hence a constant that can be evaluated at a convenient point; at $\tau = i\infty$ it is equal to $-256\pi$. The remaining integral in $\tau$ computes the area of Teichm\"uller space, equal to $\pi/3$.

\subsection{BPS States with Arbitrarily Large $\U(1)$ Charge}

We now extend the analysis to string states associated with massive, charged perturbative BPS states in three dimensions. One way to obtain such states is to consider right-moving ground states with large compact momentum $M \in\bbZ$ and winding $L$ along the $S_{R}^{1}$ with $ML<0$. These states have left moving excitation number $N_{L} = 1 - ML$ and therefore are excited string tower modes. To capture excited modes of a single heterotic string wrapping $S_{R}^{1}$ we may take $L = 1$, but our setup allows for more general combinations of $M$ and $L$. There is a large degeneracy of string states associated with a given value of $-ML$; we focus on an infinite class of states of variable $\U(1)$ charge for which computation is straightforward. This class of states is characterized by having the $-ML$ excitations spread out among the lowest lying states of a mode $\lambda^{\bar{\imath}}$ in the sigma-model $\mcC$. Such states arise in the NS-NS sector of the heterotic string and are given by
\[\ket{N} = \lambda^{\bar{\imath}}_{-N-1/2}\lambda^{\bar{\imath}}_{-N+1/2}\dots \lambda^{\bar{\imath}}_{-3/2}\lambda^{\bar{\imath}}_{-1/2}| \tilde k\rangle_{L}\otimes \psi^{i}_{-1/2}\ket{k}_{R},\qquad N_{L} = -ML \in \bbN,\addtag\label{eq:infclass}\]
where $i$ is a fixed $\SU(3)$ index and is not summed over. We consider states that survive both the GSO and orbifold projections, which puts a constraint on the values of $N$ that are allowed. GSO projection demands $(-1)^{F_{L}} = 1$, so $N$ must be odd. Invariance under $\bbZ_{3}$ transformations demands $N \in 3\bbZ$. To satisfy both conditions simultaneously we take $N \in 6\bbZ+3$. The $\U(1)$ charge $Q$ of these states is equal to $N+1$. As we will soon discover, the one-loop mass correction for these states is proportional to $Q$, so for states of large $N$ an apparent breakdown of perturbation theory occurs when their tree-level masses receive large quantum corrections. We will leave further discussion of this peculiarity to the end of this section.

Our strategy to evaluate the one-loop mass correction for the states \eqref{eq:infclass} is to extract the coefficient $F_{1}^{\qty({a'\brack b'}, {\mathbf{c} \brack \mathbf{d}})}$ defined in \eqref{eq:laur} and apply equation \eqref{eq:oneloop}. The first step is to write down the vertex operators associated with these states. Since these states are in the right-moving ground state they share the same right-moving vertex operator $V_{R}$ as the massless state \eqref{eq:mlscalar}. Using a mode expansion of NS-sector fermions, one finds that the left-moving vertex operator $V_{L}^{(Q=N+1)}$ associated to the state $\ket{N}$ is given by

\[V_{L}^{(Q)\bar{\imath}}(\tilde z) = e^{i\tilde{k}\cdot \tilde{X}} O^{(\tilde{h},Q)\bar{\imath}}(\tilde{z}),\qquad O^{(\tilde{h},Q)\bar{\imath}}(\tilde{z}) = \frac{i}{ \prod_{s=0}^{N} s!} \partial^{N}\lambda^{\bar{\imath}}(\tilde z)\partial^{N-1}\lambda^{\bar{\imath}}(\tilde z) \dots \partial\lambda^{\bar{\imath}}(\tilde z) \lambda^{\bar{\imath}}(\tilde z)\addtag\]
where $\bar\imath$ is the same index as that appearing in \eqref{eq:infclass}; for brevity we will suppress it for the remainder of this section. The prefactor is inserted to ensure that $V_{L}^{(Q)}$ creates when acting on the left-moving vacuum a state of unit norm. The vertex operator associated to the state \eqref{eq:infclass} is therefore $V^{(Q)}(\tilde{z}, z) = V_{L}^{(Q)}(\tilde{z})V_{R}(z)$. Fixing a twist $(\mathbf{c}, \mathbf{d})$, and left-moving spin structure $(a',b')$, we compute the correlator $\left\langle O^{(\tilde{h}, Q)}(\tilde{z}_{1})\bar{O}^{(\tilde{h}, -Q)}(\tilde{z}_{2}) \right\rangle_{T^{2}}$ using Wick's theorem. A quick counting exercise reveals that there are $(N+1)!$ distinct contractions, which come about in the following way. Consider the ``diagonal term'' where $\partial^{n}\lambda(\tilde z_{1})$ contracts with $\partial^{n}\lambda(\tilde z_{2})$ for each $0\leq n\leq N$:

\[ \wick{(\partial^{N} \c1\lambda)\dots (\partial^{2}\c2\lambda)(\partial\c3\lambda)\c4\lambda(\tilde z_{1})\hspace{2em}(\partial^{N} \c1\lambda)\dots (\partial^{2}\c2\lambda)(\partial\c3\lambda)\c4\lambda(\tilde z_{2})} \addtag \quad = \left\langle O^{(\tilde{h}, Q)}(\tilde{z}_{1})\bar{O}^{(\tilde{h}, -Q)}(\tilde{z}_{2}) \right\rangle^{\qty({a'\brack b'},{\mathbf{c}\brack \mathbf{d}})}_{T^{2}, \text{ id}}\label{eq:diagonalterm}\]
Starting with the diagonal term, we can obtain any other term in $\left\langle O^{(\tilde{h}, Q)}(\tilde{z}_{1})\bar{O}^{(\tilde{h}, -Q)}(\tilde{z}_{2}) \right\rangle_{T^{2}}$ by permuting the right legs of the contractions. If we associate the order $n$ of the derivative acting on $\lambda$ with the factor $\partial^{n}\lambda$ in $O$, the various permutations $ \left\langle O^{(\tilde{h}, Q)}(\tilde{z}_{1})\bar{O}^{(\tilde{h}, -Q)}(\tilde{z}_{2}) \right\rangle_{T^{2}, \sigma}$ can be identified with elements $\sigma$ of the symmetric group $S_{N+1}$ acting on the set $I = \{0,1,\dots, N\}$ (note the unusual labeling of the representation). Thus, in this setup the diagonal term is associated with the identity permutation. To illustrate this, we list some permutations in $S_{N+1}$ and their associated contractions:

\[\begin{aligned}
(10) & \quad\leftrightarrow\quad \wick{(\partial^{N} \c1\lambda)\dots (\partial^{2}\c2\lambda)(\partial\c3\lambda)\c4\lambda(z_{1})\hspace{2em}(\partial^{N} \c1\lambda)\dots (\partial^{2}\c2\lambda)(\partial\c4\lambda)\c3\lambda(z_{2})} \quad = \left\langle O^{(\tilde{h}, Q)}(\tilde{z}_{1})\bar{O}^{(\tilde{h}, -Q)}(\tilde{z}_{2}) \right\rangle^{\qty({a'\brack b'},{\mathbf{c}\brack \mathbf{d}})}_{T^{2}, (10)} \\
(21) & \quad\leftrightarrow\quad \wick{(\partial^{N} \c1\lambda)\dots (\partial^{2}\c2\lambda)(\partial\c3\lambda)\c4\lambda(z_{1})\hspace{2em}(\partial^{N} \c1\lambda)\dots (\partial^{2}\c3\lambda)(\partial\c2\lambda)\c4\lambda(z_{2})} \quad = \left\langle O^{(\tilde{h}, Q)}(\tilde{z}_{1})\bar{O}^{(\tilde{h}, -Q)}(\tilde{z}_{2}) \right\rangle^{\qty({a'\brack b'},{\mathbf{c}\brack \mathbf{d}})}_{T^{2}, (21)}\\
(210) & \quad\leftrightarrow\quad \wick{(\partial^{N} \c1\lambda)\dots (\partial^{2}\c2\lambda)(\partial\c3\lambda)\c4\lambda(z_{1})\hspace{2em}(\partial^{N} \c1\lambda)\dots (\partial^{2}\c3\lambda)(\partial\c4\lambda)\c2\lambda(z_{2})} \quad = \left\langle O^{(\tilde{h}, Q)}(\tilde{z}_{1})\bar{O}^{(\tilde{h}, -Q)}(\tilde{z}_{2}) \right\rangle^{\qty({a'\brack b'},{\mathbf{c}\brack \mathbf{d}})}_{T^{2}, (210)}\\
\vdots \hspace{1em} &\hspace{15.2em}\vdots \hspace{20em} \vdots
\end{aligned}\addtag\label{eq:sgact}\]
By spacetime statistics, permuting the legs of a contraction comes with a cost. A leg moved one fermion over incurs a factor of $-1$, and so the total product of minus signs of each permuted contraction, relative to the diagonal term, can be identified as the sign $\sgn(\sigma)$ of the permutation $\sigma \in S_{N+1}$ to which the term is associated. Additionally, the diagonal term itself has a number of minus signs equal to $(-1)^{N(N+1)/2}$ coming from permuting fermions while performing contractions. In summary,

\[ \left\langle O^{(\tilde{h}, Q)}(\tilde{z}_{1})\bar{O}^{(\tilde{h}, -Q)}(\tilde{z}_{2}) \right\rangle_{T^{2}}^{\qty({a'\brack b'}, {\mathbf{c}\brack \mathbf{d}})} = (-1)^{N(N+1)/2} \sum_{\sigma\in S_{N+1}} \sgn(\sigma) \left\langle O^{(\tilde{h}, Q)}(\tilde{z}_{1})\bar{O}^{(\tilde{h}, -Q)}(\tilde{z}_{2}) \right\rangle^{\qty({a'\brack b'},{\mathbf{c}\brack \mathbf{d}})}_{T^{2}, \sigma}.\addtag\label{eq:sgcont}\]

For a given orbifold twist $(\mathbf{c}, \mathbf{d})$ and left-moving spin $(a', b')$ structure, consider the correlator $M_{i,j}(\tilde z_{1} - \tilde z_{2}) = (1/i!j!) \left\langle \partial^{i}\lambda(\tilde z_{1}) \partial^{j}\lambda(\tilde z_{2}) \right\rangle$ in that sector. It can be evaluated using standard expressions for fermion propagators, which we expand in the on-shell limit:
\[M_{i,j}(\tilde z_{1}-\tilde z_{2}) = \frac{1}{i!j! \theta{a' + c_{1}\brack b' + d_{1}}(0)} \sum_{n=0}^{i+j} \sum_{m=0}^{n} \sum_{u=0}^{\infty} \frac{(-1)^{n}n!}{[\theta'_{1}(0)]^{m+1}} \binom{i+j}{n} \theta^{(i+j+u-n)}{a' + c_{1}\brack b' +d_{1}}(0) \frac{B_{n,m}(0)}{(\tilde z_{1}-\tilde z_{2})^{m-u+1}}.\addtag\]
Here $B_{n,m}(\tilde z_{1}-\tilde z_{2})$ are the Bell polynomials in the $m-n+1$ variables $\theta_{1}'(\tilde z_{1}-\tilde z_{2}), \dots, \theta^{(m-n+1)}(\tilde z_{1}-\tilde z_{2})$, defined for $0\leq n \leq m$. Using equation \eqref{eq:sgcont} we can evaluate the correlator in this limit,
\begin{multline}
\left\langle O^{(\tilde{h}, Q)}(\tilde{z}_{1})\bar{O}^{(\tilde{h}, -Q)}(\tilde{z}_{2}) \right\rangle^{\qty({a'\brack b'}, {\mathbf{c}\brack \mathbf{d}})}_{T^{2}} = (-1)^{N(N+1)/2} \sum_{\sigma \in S_{n+1}}\sgn(\sigma) \prod_{i=0}^{N} M_{i,\sigma(i)} (\tilde{z}_{1}-\tilde{z}_{2}) \\
= \frac{(-1)^{N(N+1)/2}}{\theta^{N+1}{a +c_{1}\brack b + d_{1}}(0)} \sum_{\sigma\in S_{N+1}} \sgn(\sigma) \prod_{i=0}^{N} \frac{1}{i!\sigma(i)!} \sum_{n_{i}=0}^{i+\sigma(i)} \sum_{m_{i} = 0}^{n_{i}} \sum_{u_{i}=0}^{\infty}\\
\qty{ \frac{(-1)^{n_{i}}n_{i}!}{[\theta'_{1}(0)]^{m_{i}+1}} \binom{i+\sigma(i)}{n_{i}} \theta^{(i+\sigma(i)+u_{i}-n_{i})}{a' + c_{1}\brack b' +d_{1}}(0) \frac{B_{n_{i},m_{i}}(0)}{(\tilde z_{1}-\tilde z_{2})^{m_{i}-u_{i}+1}}}
\end{multline}
To extract the first subleading terms whose overall coefficient is $F_{1}^{\qty({a' \brack b'}, {\mathbf{c}\brack \mathbf{d}})}$, we focus on the powers of $\tilde z_{1}-\tilde z_{2}$ in the expansion. The on-shell condition $\tilde{k}^{2} + 2N_{L} - 2 = k^{2}$ leads to the following selection rule on the indices: $\sum_{i=0}^{N}(u_{i} - m_{i}) + 2N_{L} - 1 = 0$. Using the fact that $N_{L} = \sum_{i=0}^{N} i = \sum_{i=0}^{N}\sigma(i)$ for all $\sigma\in S_{N+1}$, we write this suggestively as
\[ \sum_{i=0}^{N} u_{i} + \sum_{i=0}^{N}(i+\sigma(i) - n_{i}) + \sum_{i=0}^{N}(n_{i} - m_{i}) = 1.\addtag\]
Observe that each summand in this equation is non-negative. Therefore, the contributions to $F_{1}^{\qty({a' \brack b'}, {\mathbf{c}\brack \mathbf{d}})}$ are in correspondence with the choices of whichever term is equal to one. There are three types of terms:
\begin{enumerate}
\item those with $u_{i} = 1$ for some $i\in I$, $u_{j} = 0$ for all $j\neq i$, and $n_{i} = m_{i} = i + \sigma(i)$ for all $i \in I$;
\item those with $n_{i} = m_{i} = i + \sigma(i) - 1$, for some $i\in I$, $n_{j} = m_{j} = j + \sigma(j)$ for all $j\neq i$, and $u_{i} = 0$ for all $i \in I$; and
\item those with $n_{i} = m_{i} + 1 = i + \sigma(i)$ for some $i\in I$, $n_{j} = m_{j} = j + \sigma(j)$ for all $j\neq i$, and $u_{i} = 0$ for all $i \in I$.
\end{enumerate}
The terms of type $(2)$ and $(3)$ cancel exactly due to the factor of $(-1)^{n_{i}}$ in the sum. The terms of type (1) are all identical and there are $N+1$ of them. Thus, we find that
\[F_{1}^{\qty({a' \brack b'}, {\mathbf{c}\brack \mathbf{d}})} = (N+1)\times \frac{\theta'{a' + c_{1}\brack b' + d_{1}}(0)}{\theta{a' + c_{1}\brack b' + d_{1}}(0)} \sum_{\sigma\in S_{N+1}} \sgn(\sigma) \prod_{i=0}^{N} \binom{i+\sigma(i)}{i} = (N+1) \times \frac{\theta'{a' + c_{1}\brack b' + d_{1}}(0)}{\theta{a' + c_{1}\brack b' + d_{1}}(0)}\times \det L^{(N)}.\addtag\label{eq:F1m}\]
In equation \eqref{eq:F1m}, $L^{(N)}$ is an $(N+1)$-by-$(N+1)$ matrix with matrix elements $L^{(N)}_{i,j} = \binom{i+j}{i}$. We will shortly prove that $\det L^{(N)} = 1$ for all values of $N$, but with this result we can immediately apply equation \eqref{eq:oneloop} and use the integral identity \eqref{eq:inttau} to arrive at the conclusion that the one-loop mass renormalization of this class of charge-$(N+1)$ massive BPS states is equal to
\[m^{2}_{\text{one-loop}} = 12(N+1)g_{s}^{2}.\addtag\]

It remains to prove the claim that the determinant of the matrix $L^{(N)}$ is equal to $1$ for all values of $N$. We illustrate our strategy with an example in low rank. The basic idea is to use column operations, which leave the determinant invariant, to inductively relate $\det L^{(N)}$ and $\det L^{(N-1)}$. For the case $N=2$ (so $L$ is $3$-by-$3$), the desired operation is to add $1$ times the first column and $-2$ times the second column to the last column:
\[L^{(2)} = \begin{pmatrix}
1 & 1 & 1 \\ 1 & 2 & 3 \\ 1 & 3 & 6
\end{pmatrix} \qquad \to \qquad L^{(2),\text{aug}} = \begin{pmatrix}
1 & 1 & 0 \\ 1 & 2 & 0 \\ 1 & 3 & 1
\end{pmatrix}.\addtag\]
From this manipulation it is immediately clear that $\det L^{(2)} = 1 \times \det L^{(1)} = 1$ by the inductive hypothesis. The proof to the general statement, which we leave to Appendix \ref{sec:combid}, proceeds in this spirit.

Note that the one-loop mass renormalization of charged BPS states is proportional to their $\U(1)$ charge $Q$, a fact which has interesting consequences for superstring perturbation theory in this class of models.  For any finite value of string coupling $g_{s}$, there exists an $Q^*$ for which the one-loop correction to the mass of charged BPS states is comparable to their tree-level value, suggesting that perturbation theory becomes unreliable. For states with $L=1$ whose mass is dominated by the winding contribution this occurs roughly at
\[ Q^* \gtrsim \frac{R^{2}}{g_{s}^{2}\alpha'}.\addtag\]

\subsection{An Approach via Super Riemann Surfaces}\label{ssec:susyform}

In this section, we offer an alternative derivation of the results obtained above, using the formalism of perturbation theory on super-Riemann surfaces \cite{Witten:2012bh,Witten:2012ga, Witten:2019uux}. An attractive aspect of this viewpoint is that we are able to take $\ket{V}$ to be on-shell at the beginning of the computation, avoiding conceptual complications associated with off-shell dynamics, whose adequate treatment requires an understanding of string field theory. For ease of comparison to the references listed above, we adopt the same convention which assigns anti-holomorphic coordinates $\tilde{z}$ to the left-moving bosonic string and holomorphic supercoordinates $z\vert\theta$ to the right-moving superstring. It is important that the world-sheet coordinate $\tilde{z}$ not be identified with the complex conjugate of the coordinate $\bar{z}$ and the choice of notation reflects this. In this formalism, target space coordinates for $\IR^{2,1}$ and the Calabi-Yau threefold $\mcY$ are paired with their supersymmetric partners
\[\begin{aligned}
\mathscr{X}^{\mu}(\tilde{z}, z\vert \theta) = X^{\mu}(\tilde{z}, z) + \theta \psi^{\mu}(\tilde{z}, z),\qquad \mathscr{Y}^{i}(\tilde{z}, z\vert \theta) = Y^{i}(\tilde{z}, z) + \theta \psi^{i}(\tilde{z}, z).
\end{aligned}\addtag\]
For the coordinate on $S^1_R$ we must also decompose the coordinates (and momenta) into left and right-moving components
\begin{equation}
\mathscr{X}^{3}(\tilde{z}, z\vert \theta) = \tilde X^{3}(\tilde{z}) + X^{3}(z) + \theta \psi^{3}(z).
\end{equation}
Additionally, in the fermionic formulation of the heterotic string, there are fermions $\lambda^{T}$ with $\SO(26)$ index (here we use $T$ instead of $P$, conforming to the conventions of \cite{Witten:2012bh,Witten:2012ga,Witten:2019uux}), which we pair up with bosonic auxiliary fields $G^{T}(\tilde{z}, z)$ that vanish on-shell
\[\Lambda^{T}(\tilde{z}, z\vert \theta) = \lambda^{T}(\tilde{z}, z) + \theta G^{T}(\tilde{z}, z).\addtag\]
and similarly for $\Lambda^i$ and $\lambda^i$.
Using the superspace derivative $D_{\theta} = \partial_{\theta} + \theta \partial_{z}$, one can form the superfield vertex operators associated with the infinite class of massive BPS states we have considered thus far
\[W^{(Q)}_{\tilde k, k}(\tilde{z}, z\vert \theta) = \mcO^{(\tilde{h} -1/2, Q-1)}_{\text{aux}} (\tilde{z}, z\vert\theta)e^{i \tilde{k} \cdot \tilde{\mathscr{X}}} \Lambda_i D_{\theta}\mathscr{Y}^{i}(z\vert\theta) e^{ik\cdot \mathscr{X}},\addtag\]
and its CPT conjugate $\bar W^{(-Q)}_{-\tilde k, -k}$.  
Here  $\mcO_{\text{aux}}^{(\tilde{h}-1/2,Q-1)}$ is a world-sheet CFT operator which reduces on shell to $O_{\text{aux}}^{(\tilde{h}-1/2,Q-1)}$, defined below equation \eqref{eq:JOope}. The two-point function we must evaluate then takes the form
\[m^{2}_{\text{one-loop}} = -g_{s}^{2} \int \frac{\dd[2]{\tau}}{4\Im\tau} \int \dd[2]{z} \dd{\theta}\dd{\theta'} \left\langle W^{(Q)}_{\tilde k, k}(\tilde{z}, z\vert \theta) \bar W^{(-Q)}_{-\tilde k, -k}(0,0\vert \theta') \right\rangle,\addtag\]
where we have used translation invariance on the torus to move one of the operator insertions to $\tilde{z} = z = 0$. As explained in \cite{Witten:2019uux}, although the OPE of $W$ with the conjugate operator $\bar W$ contains many terms, the only contribution to the integral arises as before from the operator $V_{D}$ of dimension $(1,1)$, which can
be thought of as the vertex operator for the auxiliary $D$ field in the $\U(1)$ gauge supermultiplet.  In the computation of the OPE and integral it is possible to work directly
on mass-shell. In \cite{Witten:2019uux} where this method was used to compute the mass correction to massless states with $k^2=0$, it was possible to simply take $k^\mu=0$.
Here this is not possible because we are dealing with massive states, but it is still possible to work on mass-shell where $k \ne 0$ but $k^2=0$. Computing
the coefficient of $V_D$ in the OPE at small $\tilde z, z$  leads to
\begin{equation}
W^{(Q)}_{\tilde k, k}(\tilde{z}, z\vert \theta) \bar W^{(-Q)}_{-\tilde k, - k}(0,0\vert \theta')  \sim \frac{V_D(0;0)}{\tilde z^{2\tilde{h}- 1+ \tilde k^2} z^{k^2}}
\end{equation}
but on-shell we have $2\tilde{h} - 1 + \tilde k^2= k^2+1=1$ so we are left with a result proportional to $\langle V_D \rangle$ and the integral
\begin{equation}
\int d^2 z d \theta d \theta' \frac{1}{\tilde z} \, .
\end{equation}
There are subtleties with carrying out this integral because it was obtained only near $z=\tilde z=0$ and at large $\tilde z$ it must be modified to deal with the
periodic nature of the torus. However these subtleties are dealt with in full detail in  \cite{Witten:2019uux} and do not need to repeated here. The integral
has the value $2 \pi$, and when combined with the other factors we have suppressed replicates the result we found earlier by going on-shell only at the
end of the computation. 

%
%

\section{Conclusions and Outlook}

In this work we have studied the one-loop correction to the mass of perturbative BPS states when the $\SO(32)$ heterotic string is compactified on $\mcY \times S^1_R$
where $\mcY$ is a Calabi-Yau threefold with $\chi(\mcY) \ne 0$ and the gauge symmetry is broken to $\SO(26) \times \U(1)$ or $\SO(26) \times \SU(3) \times \U(1)$ in the orbifold limit.
The one-loop correction is of order $g_s^2$ and proportional to their $\U(1)$ charge. These BPS states have non-zero winding number on the $S^1$ factor and can be viewed as either BPS string states in four spacetime dimensions at large $R$ or as BPS states in three dimensions at small $R$. The one-loop correction we have found arises
in a way that is closely related to the mass correction for massless states computed long ago \cite{Atick:1987gy, Dine:1987gj}. We also showed that the super Riemann surface
formalism that was applied to the massless state mass correction in \cite{Witten:2019uux} can be extended to compute the BPS mass correction and avoids the issue
of having to work slightly off-shell.

The BPS mass corrections we have computed give the masses in the unshifted vacuum, where the scalars have not developed vacuum expectation values. In the shifted vacuum, where the D-terms vanish and supersymmetry is restored, there are additional terms which can modify the BPS masses. For example, there can
be terms in the scalar potential of the form $(\Phi^{(Q)})^\dagger \Phi^{(Q)} \phi^\dagger \phi$ where $\Phi^{(Q)}$ is a charge $Q$ field creating a BPS state and $\phi$ is one of the charge $-2$ massless scalars with a non-zero
vacuum expectation value. Since $\langle \phi^\dagger \phi \rangle$ is also order $g_s^2$, these terms give additional mass corrections of the same order in $g_s$ as
the terms we have computed. However these couplings are typically independent of the charge $Q$ and so at least at large $Q$ the terms we have computed should
give the leading correction to the masses of BPS states. 

There are several interesting questions that we hope to pursue in the near future. The first is whether the direct mass correction to BPS states is purely a one-loop effect or whether there are corrections at higher order in string perturbation theory. In the low-energy limit of supersymmetric quantum field theory there are arguments that
the induced FI-like terms related to the mass correction appear only at one-loop \cite{Fischler:1981zk} but to our knowledge these arguments have not been extended to
string perturbation theory.  A full computation of the BPS mass spectrum also involves a  study of their masses in the shifted vacuum due to their couplings to the scalar field
with a supersymmetry restoring vacuum expectation value. This seems like a doable, but complicated computation. 
 A second question is that of non-perturbative corrections to BPS masses arising from instanton effects. If it is correct that perturbative corrections arise
only at one-loop then the situation is somewhat reminiscent of Seiberg-Witten theory and, as there, one might expect that non-perturbative effects play a crucial role
in providing a consistent picture of BPS masses.  Third, we have computed a perturbative correction to the mass of BPS states and normally one would expect to view this as a perturbative correction to the central charge. It is not entirely clear to us if that is true here as well due to the subtleties in defining asymptotic supercharges in three dimensions.
Finally, this project started with a study of the hybrid axion-gauge cosmic strings which appear in string compactifications
with pseudo-anomalous $\U(1)$ factors. It has been known for some time that the massless spacetime fields outside these strings become singular near the core
of the string \cite{Binetruy:1998mn} and we believe that the computations done here will help to provide an understanding of the origin of these singularities \cite{inprep}. 

\section*{Acknowledgements}

We thank Michael Levin for assistance with the identity proved in Appendix A and  Clay Cordova, Atish Dabholkar, Greg Moore, and Ashoke Sen for helpful comments. This work was supported by the National Science Foundation Grant PHY-2310635.

\appendix
\section{Proof of Determinant Identity}\label{sec:combid}

In this appendix, we prove by induction the statement that the matrix $L^{(N)}$ whose matrix elements are $L^{(N)}_{i,j} = \binom{i+j}{i}$ has determinant one. Written out explicitly, $L^{(N)}$ takes the form

\[L^{(N)} = \begin{pmatrix}
1 & 1 & 1 & 1 & \dots & \binom{N}{0} \\
1 & 2 & 3 & 4 & \dots & \binom{N+1}{1} \\
1 & 3 & 6 & 10 & \dots & \binom{N+2}{2} \\
1 & 4 & 10 & 20 & \dots & \binom{N+3}{3} \\
\vdots & \vdots & \vdots & \vdots & \ddots & \vdots \\
\binom{N}{0} & \binom{N+1}{1} & \binom{N+2}{2} & \binom{N+3}{3} & \dots & \binom{2N}{N}
\end{pmatrix}.\addtag\]
To simplify some expressions, we count rows and columns starting from zero. Note that this is different from the usual indexing of matrix elements, but is convenient given the unorthodox form of symmetric group action \eqref{eq:sgact}. The base case with $N=0$ and $L^{(0)} = (1)$ is obviously true. For the inductive step, we relate $L^{(N)}$ and $L^{(N-1)}$ using linear column operations. We add $(-1)^{N+j}\binom{N}{j}$ times the $j$th column to the last column for each $j=0,\dots, N-1$:

\[L^{(N)}_{-,N} \quad\to\quad L^{(N)\text{,aug}}_{-,N} = L^{(N)}_{-,N} + \sum_{j=0}^{N-1} (-1)^{N+j} \binom{N}{j} L^{(N)}_{-,j}.\addtag\]
Linear column and row operations leave the determinant invariant, so it suffices to compute the determinant of the augmented matrix. Let's see the effect of this manipulation on the last column of $L^{(N)}$. We look row-by-row. The last entry of the $i$th row in $L^{(N), \text{aug}}$ is equal to

\[\binom{N+i}{i} + \sum_{j=0}^{N-1} (-1)^{N-j} \binom{n}{j}\binom{i+j}{j} = (-1)^{N} \sum_{j=0}^{N} (-1)^{j} \binom{N}{j} \binom{i+j}{j} = (-1)^{N} \sum_{j=0}^{N} \binom{N}{j} \binom{-i-1}{j}.\addtag\]
To proceed, we find the following combinatorial identity useful:

\[ \sum_{j=0}^{n} \binom{n}{j} \binom{m}{j} = \binom{n+m}{n}.\addtag\]
It can be understood by the following double counting argument. Suppose there are $m$ people in group A and $n$ people in group B. How many committees of $n$ people can we form? On one hand, it is $\binom{n+m}{n}$. On the other, we can select $j$ people from group $B$ and $n-j$ people from group A, and sum up the possibilities for each $0\leq j\leq n$:

\[\binom{n+m}{m} = \sum_{j=0}^{n} \binom{n}{n-j} \binom{m}{j} = \sum_{j=0}^{n} \binom{n}{j} \binom{m}{j}.\addtag\]
Applying this identity to the above with $m= -i -1$ and $n=N$, we find the last entry of the $i$ row of the augmented matrix to be equal to

\[L^{(N)\text{,aug}}_{i,N} = (-1)^{N} \sum_{j=0}^{N}\binom{N}{j} \binom{-i-1}{j} = (-1)^{N} \binom{N-i-1}{N}.\addtag\]
For $0\leq i\leq N-1$, $0\leq N-i-1 < N$ and $L^{(N)\text{, aug}}_{i,N} = 0$. However, for $i=N$ (the lower right-most entry of $L^{(N)}$)

\[L^{(N)\text{,aug}}_{N,N} = (-1)^{N}\binom{-1}{N} = \binom{N}{N}=1.\addtag\]
It follows that $\det L^{(N)} = \det L^{(N),\text{aug}} = 1\times \det L^{(N-1)} = \det L^{(N-1)}$, and we are done.

\end{document}